# Near-complete violation of Kirchhoff's law in thermal radiation in ultrathin magnetic Weyl semimetal films

Jun Wu[1], Zhongmin Wang[2], Han Zhai[3], Zhangxing Shi[3], Xiaohu Wu[3,*], and Feng Wu[4,*]

1. Key Laboratory of Advanced Perception and Intelligent Control of High-end Equipment, Ministry of Education, College of Electrical Engineering, Anhui Polytechnic University, Wuhu, 241000, China

2. Institute of Automation, Qilu University of Technology (Shandong Academy of Sciences), Jinan 250014, China

3. Shandong Institute of Advanced Technology, Jinan 250100, China

4. School of Optoelectronic Engineering, Guangdong Polytechnic Normal University, Guangzhou 510665, China

*Corresponding author: xiaohu.wu@iat.cn, fengwu@gpnu.edu.cn

**Abstract:** The ability to break Kirchhoff's law is of fundamental importance in thermal radiation. Various nonreciprocal emitters have been proposed to break the balance between absorption and emission. However, the thicknesses of the nonreciprocal materials are usually larger than 1/10 times of the wavelength. Besides, the previous proposed nonreciprocal emitters are complex, thus they can hardly be fabricated in experiment to verify the Kirchhoff's law for nonreciprocal materials. In this paper, we investigate the nonreciprocal thermal radiation of the magnetic Weyl semimetal (MWSM) film atop of the metal substrate. It is found that the strong nonreciprocal radiation at the wavelength of 9.15 μm can be achieved when the thickness of the MWSM film is 100 nm. The enhanced nonreciprocity is attributed to the Fabry-Perot resonances. The results indicate that the MWSM film is the promising candidate to engineer the ultrathin and simple nonreciprocal thermal emitters. What is perhaps most intriguing here is that the proposed structure can be more easily fabricated in experiment to verify the Kirchhoff's law for nonreciprocal materials.

**Keywords:** nonreciprocal thermal radiation, magnetic Weyl semimetal, Fabry-Perot resonances, Kirchhoff's law

# 1. Introduction

All objects at nonzero temperatures emit electromagnetic radiation according to the fundamental principle of statistical mechanics [1-4]. It has been long recognized in the literature that most thermal emitters obey the Kirchhoff's emission-absorption equivalence law. Therefore, the Kirchhoff's law plays a very significant role in understanding thermal emission. The recent emergence of nonreciprocal thermal emitters has opened new avenues for controlling fundamental aspects of thermal emission, because such emitters can break the Kirchhoff's law, i.e., breaking the balance between absorption and emission [5-23]. Since the traditional Kirchhoff's law is not applicable for nonreciprocal thermal emitters, the generalized Kirchhoff's law that is applicable for both reciprocal and nonreciprocal thermal emitters has been explored and derived [24-27].

Various nonreciprocal materials, including magneto-optical materials and magnetic Weyl semimetals (MWSMs), are engineered to break the equivalence between absorption and emission [11-23, 28]. Compared with magneto-optical materials, MWSMs possess intrinsic advantages, i.e., it has stronger nonreciprocity in the mid-infrared range without applying an external magnetic field [20-23]. Therefore, the MWSMs are the promising candidates for nonreciprocal thermal emitters. Besides, nonreciprocal thermal emitters containing magneto-optical materials usually need large thickness to enhance the difference between absorption and emission [11-17]. What is most important, complex structures are usually needed [11-17]. Although MWSMs possess intrinsic nonreciprocity, they have not been comprehensively investigated to engineer nonreciprocal thermal emitters, especially for ultrathin and simple nonreciprocal thermal emitters. Grating structures and prism-coupling structures have been used to enhance the difference between absorption and emission of MWSMs [20, 23]. Due to the fabrication technologies, these structures are not easy to be fabricated in experiment at present. The fabrication of single MWSM film has been achieved by several groups [29-31]. However, the nonreciprocal thermal radiation of single MWSM film has not been fully explored.

In this work, the nonreciprocal radiation of a single MWSM film atop of the metal substrate is explored. The results show that strong nonreciprocal radiation at the wavelength of 9.15 μm can be achieved when the thickness of the MWSM film is 100 nm. The enhanced nonreciprocity is attributed to the Fabry-Perot (FP) resonances. Our results show that the MWSM film is the promising candidate to engineer the ultrathin and simple nonreciprocal thermal emitters. Besides, the structure is promising to be used to verify the Kirchhoff's law for nonreciprocal materials, since it can be easily fabricated in experiment [31].

## 2. Model

In this work, the proposed structure is shown in Fig. 1, where a MWSM film with a thickness of $d_1$ is on the top of the silver (Ag) substrate. The relative permittivity of Ag is described by the Drude model, i.e., $\varepsilon_{Ag} = \varepsilon_\infty - \omega_p^2/(\omega^2 + j\omega\Gamma)$, with $\varepsilon_\infty = 3.4$, $\omega_p = 1.39 \times 10^{16}$ rad/s, $\Gamma = 2.7 \times 10^{13}$ rad/s, and $\omega$ is the angular frequency [32]. The wavevector component along the x-axis is $k_x = k_0 \sin\theta$, where $k_0$ is the wavevector in the vacuum and $\theta$ is the angle of incidence.

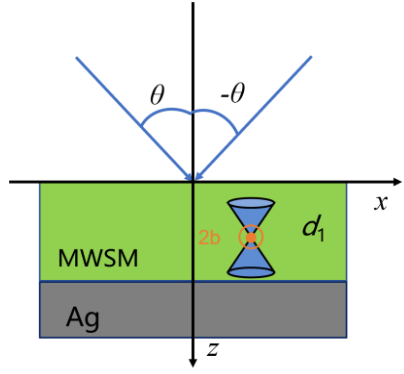

Fig. 1 Schematic of the proposed nonreciprocal thermal emitters. The MWSM film with a thickness of $d_1$ is on the top of the metal substrate.

When each pair of Weyl nodes are separated along the $y$-axis in the momentum space, the relative permittivity tensor of the MWSM can be described as [20, 23]

$$\boldsymbol{\varepsilon} = \begin{bmatrix} \varepsilon_d & 0 & -j\varepsilon_a \\ 0 & \varepsilon_d & 0 \\ j\varepsilon_a & 0 & \varepsilon_d \end{bmatrix}. \quad (1)$$

The detail expressions of the relative permittivity components can be found in Refs. [21] and [24]. When $\varepsilon_a$ is zero, the permittivity tensor is symmetric, thus obeying Lorentz reciprocity [35]. When $\varepsilon_a$ is nonzero, the permittivity tensor is asymmetric, thus breaking Lorentz reciprocity [35]. When the temperature is 300 K, two components $\varepsilon_d$ and $\varepsilon_a$ are shown in Fig. 2. $\varepsilon_d$ is a complex number while $\varepsilon_a$ is a real number. One can see that the epsilon-near-zero wavelength is about 8.6 μm. The real part $\mathrm{Re}(\varepsilon_d)$ is negative when the wavelength is larger than 8.6 μm. The $\varepsilon_a$ increases with the wavelength, indicating that stronger nonreciprocity can be achieved at larger wavelengths.

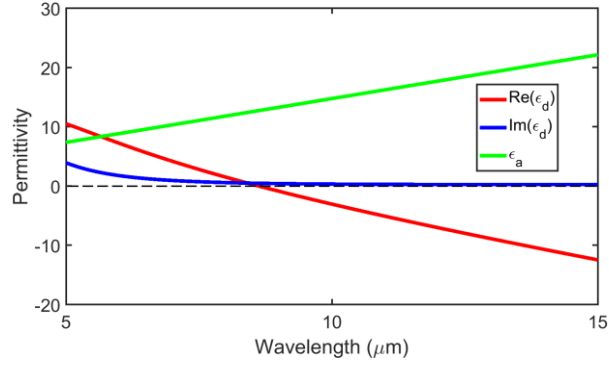

Fig. 2 Relative permittivity of the MWSM as a function of the wavelength. Black dashed line represents zero permittivity.

The plane of incidence is *x-z* plane, thus there is no polarization conversion between two linearly polarized waves. When a TM- (transverse magnetic, with the magnetic field along the direction of *y*-axis) polarized plane wave is incident with an angle $\theta$, the spectral directional absorption and emission of the structure can be calculated by [26]

$$\alpha(\theta,\lambda)=1-R(\theta,\lambda), e(\theta,\lambda)=1-R(-\theta,\lambda). \qquad (2)$$

Here, $R(\theta, \lambda)$ and $R(-\theta, \lambda)$ are the reflection for the incident angle of $\theta$ and $-\theta$ at the wavelength $\lambda$, respectively. The difference between emission and absorption is defined as $\eta=|\alpha-e|$, which measuring the nonreciprocal radiation. The transfer matrix method for calculating the reflection of the multilayer structures with nonreciprocal materials is presented in Ref. [17].

## 3. Results and discussion

The difference between the absorption and emission varying with the angle of incidence and wavelength for different thicknesses of the MWSM film $d_1$ are shown in Fig. 3. When the thickness is equal to or larger than 10 μm, the ultra-strong nonreciprocal radiation is located around the wavelength of 8.3 μm and the angle of incidence is 89°. When the thickness is 1 μm, besides the wavelength of 8.3 μm, large difference between absorption and emission can take place at the wavelength around 10.6 μm, as shown in Fig. 3(d). When the thickness is equal to or smaller than 0.1 μm, as shown in Figs. 3(e) and 3(f), the absorption and emission are almost identical at the wavelength of 8.3 μm. When the thickness is 0.1 μm, the strong nonreciprocal radiation is around the wavelength of 9.15 μm. When the thickness is 0.01 μm, the nonreciprocal radiation is quite weak for all the wavelengths and the angle of incidence. To sum up, the thickness of the MWSM film has a great impact on the nonreciprocal radiation.

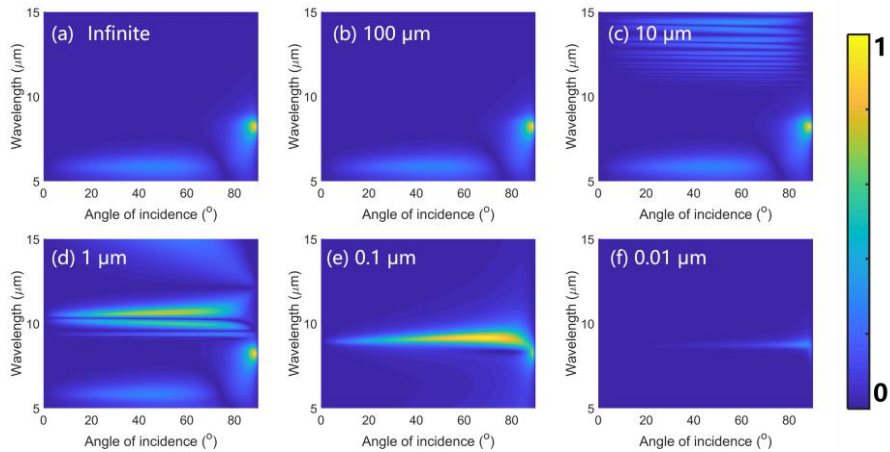

Fig. 3 Difference between the absorption and emission varying with the angle of incidence and the wavelength with difference thicknesses of the MWSM film $d_1$: (a) infinite, (b) 100 μm, (c) 10 μm, (d) 1 μm, (e) 0.1 μm, and (f) 0.01 μm.

When the thickness of the MWSM film is 10 μm and the angle of incidence is 89°, the absorption and emission are respectively shown in Fig. 4(a). One can see that the difference between the absorption and emission can reach 0.9 at the wavelength of 8.3 μm, indicating the near-complete violation of the Kirchhoff's law. The absorption and emission are almost the same when the wavelength is larger than 11 μm. In addition, the absorption and emission oscillate with the wavelength, indicating that FP resonances occur at larger wavelengths. Fig. 4(b) shows the absorption and emission varying with the thickness of the MWSM film at the wavelength of 8.3 μm when the angle of incidence is 89º. The absorption and emission do not change with the thickness when the thickness is larger than 0.4 μm, which indicates that the difference between them is stable. The absorption gets its maximum at $d_1$=73 nm with a smooth peak. The emission gets its maximum at $d_1$=27 nm with a sharper peak. These two peaks are attributed to the FP resonances. Only first order of FP resonance can be supported, either for the absorption or the emission. Such phenomenon has been observed and explained in a single hyperbolic material film [36]. To confirm that, one can analyze the wavevector in the MWSM film. The wavevector component along the z-axis in the MWSM film can be calculated by $k_{z1} = \sqrt{\varepsilon_v k_0^2 - k_x^2}$, where the effective relative permittivity is $\varepsilon_v = \varepsilon_d - \varepsilon_a^2/\varepsilon_d$. At the wavelength of 8.3 μm, we have $\varepsilon_d = 0.79 + 0.54j$ and $\varepsilon_a = 12.27$. Therefore, we have $k_{z1} = (3.72 + 11.95j)k_0$. The imaginary part of $k_{z1}$ is so large that the wave decays very fast during propagation in the film. Hence, the higher-order FP resonances cannot be supported. When the thickness of MWSM film is 10 μm, the distribution of magnetic field at the wavelength of 8.3 μm along the y-axis is plotted in Fig. 4(c). One can see that the field is strongly located at the interface between the air and the MWSM film. The field is stronger at angle of incidence of 89º than that at angle of incidence of -89º.

The stronger is the field, the smaller is the reflection. Therefore, the absorption is larger than the emission.

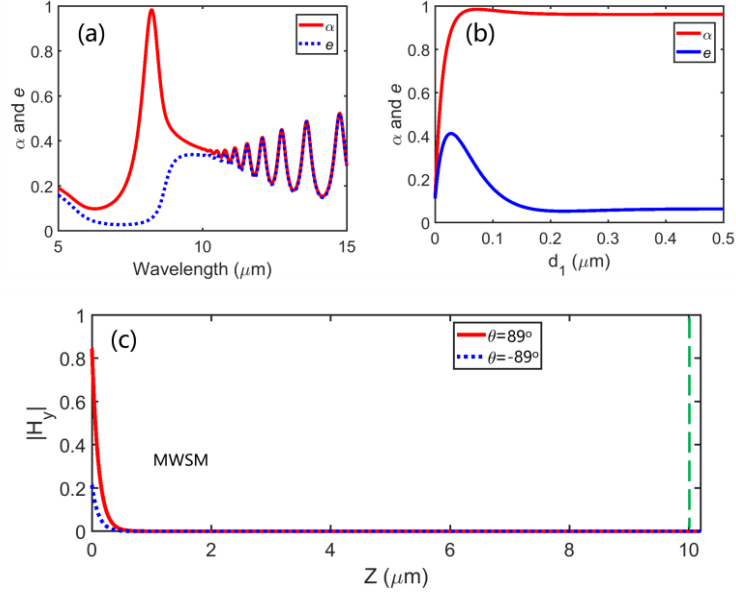

Fig. 4 (a) Absorption and emission spectra when the thickness of the MWSM film is 10 μm and the angle of incidence is 89º. (b) Absorption and emission as a function of the thickness of the MWSM film when the wavelength is 8.3 μm and the angle of incidence is 89º. (c) Distributions of the magnetic field at the wavelength of 8.3 μm at angles of incidence of 89º and -89º when the thickness of the MWSM film is 10 μm.

According to the permittivity of MWSM in Fig. 2, the wavelength of 8.3 μm is not a special point. However, the nonreciprocity at this wavelength is much larger than other wavelengths. To understand this, the reflection as functions of the angle of incidence and the wavelength is shown in Fig. 5. The thickness is infinite. Under this situation, the reflection is only related to the permittivity of the MWSM and the angle of incidence. It is clear the reflection is strong in the wavelength range between 6.5 μm and 9 μm, regardless of the sign of the angle of incidence. At the wavelength of 8.3 μm, the reflection is very small at angle of 89º, while it is large at angle of -89º.

Therefore, the nonreciprocity is strongly related with the permittivity at different wavelengths.

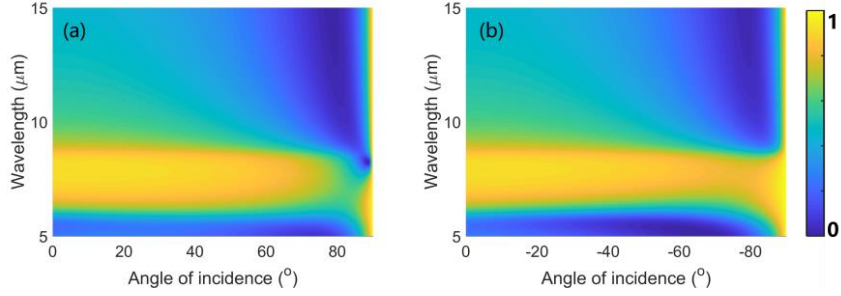

Fig. 5 Reflection as functions of the angle of incidence and the wavelength when the thickness of the MWSM film is infinite: (a) positive angle of incidence and (b) negative angle of incidence.

If $\varepsilon_a$ is zero, the reflection should be high when the wavelength is larger than 8.6 μm since $\text{Re}(\varepsilon_d)$ is negative. However, if $\varepsilon_a$ is not zero, the situation is quite different. As shown in Fig. 5, the reflection is small when the wavelength is larger than 9 μm. To understand this point, it is necessary to analyze the value of the effective relative permittivity $\varepsilon_v = \varepsilon_d - \varepsilon_a^2/\varepsilon_d$ since it impacts the wavevector in the MWSM film. The real and imaginary parts of $\varepsilon_v = \varepsilon_d - \varepsilon_a^2/\varepsilon_d$ are shown in Fig. 6. One can see that the real part of $\varepsilon_v$ is negative in the band between 5.7 μm and 8.6 μm. Besides, the peak of the imaginary part is located around 8.65 μm, and its maximum is much larger than $\text{Im}(\varepsilon_d)$ shown in Fig. 2. The negative real part of $\varepsilon_v$ leads to high reflection. Therefore, the high reflection in Fig. 5 can be understood by looking at the value of $\varepsilon_v$.

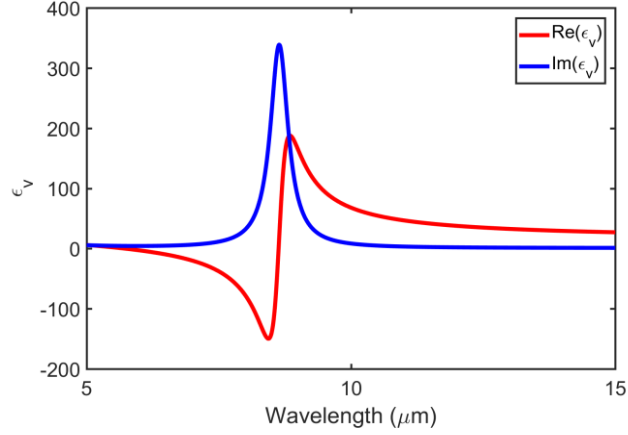

Fig. 6 Real and imaginary parts of $\varepsilon_v$ as a function of the wavelength.

When the thickness of the MWSM film is 0.1 μm and the angle of incidence is 68°, the absorption and emission are shown in Fig. 7(a). The absorption can reach 0.995 at the wavelength of 9.15 μm, while the emission is smaller than 0.03. The difference between them is higher than 0.96. Fig. 7(b) shows the absorption and emission varying with the thickness of the MWSM film. It is clear that both of them oscillate with the wavelength, indicating that FP resonances occur. The distance between adjacent peaks in the absorption spectra is about 0.39 μm, and that is also about 0.39 μm in the emission spectra. It is reasonable because the wavevectors in the MWSM film are the same, regardless of the sign of the angle of incidence. To check the excitation of FP resonances, the distance between two adjacent peaks should satisfy [36]

$$\mathrm{Re}(k_{z1})\Delta = \pi, \tag{3}$$

where $\Delta$ is the distance between two adjacent FP resonances. At the wavelength of 9.15 μm, there is $\varepsilon_d = -1.19 + 0.42j$, $\varepsilon_a = 13.52$, $k_{z1} = (11.80 + 2.05j)k_0$. According to Eq. (3), the calculated distance between two FP resonances should be 0.388 μm, which is very close to 0.39 μm. Therefore, Eq. (3) can confirm the excitation of FP resonances.

Besides, it is noted that the absorption reaches its maximum when the thickness is 0.1 μm. Therefore, the strong nonreciprocal radiation shown in Fig. 7(a) is attributed to the excitation of FP resonances in the MWSM film. The distribution of magnetic field at the wavelength of 9.15 μm along the *z*-axis is plotted in Fig. 7(b). The intensity of the incident magnetic field is set to be unity. When the angle of incidence is 68º, the magnetic field is enhanced at the interface between the MWSM film and the Ag substrate, thus the absorption is large in this case. However, the magnetic field is smaller at the interface for angle of incidence of -68º, indicating that most of the incidence wave is reflected. According to Eq. (2), the emission is small for angle of incidence of 68º.

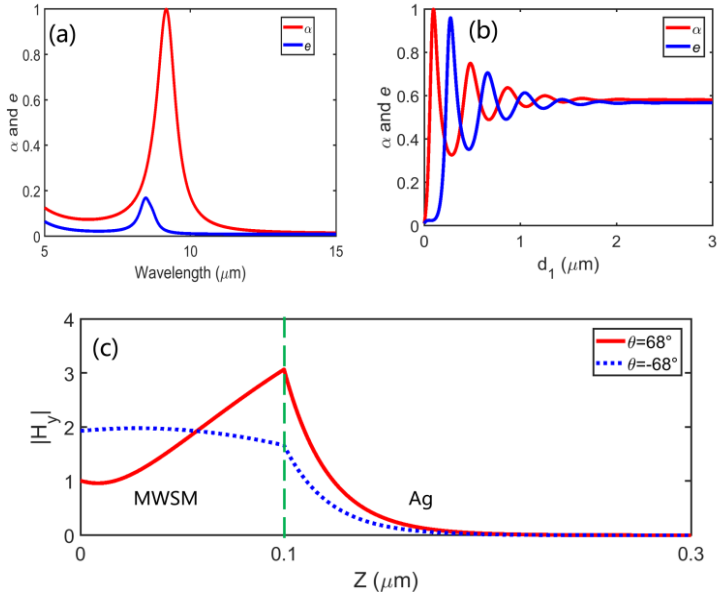

Fig. 7 (a) Absorption and emission spectra when the thickness of the MWSM film is 0.1 μm and the angle of incidence is 68º. (b) Absorption and emission as a function of the thickness of the MWSM film when the wavelength is 9.15 μm and the angle of incidence is 68º. (c) Distributions of the magnetic field at the wavelength of 9.15 μm at angles of incidence of 68º and -68º when the thickness of the MWSM film is 0.1 μm.

At the wavelength of 9.15 μm, the absorption, emission, and the difference between them as functions of the angle of incidence and the thickness of the MWSM is shown in Fig. 8. According to Figs. 8(a) and 8(b), one can see that the first order FP resonance is strong, while the other orders are weak. Besides, strong absorption and emission are realized at large angle of incidence. As shown in Fig. 8(c), it is clear the nonreciprocal radiation is strong at the first order FP resonance. In addition, the large difference between absorption and emission can be realized when the angle of incidence is larger than 40º. It is hard to realize strong nonreciprocal radiation at small angles of incidence.

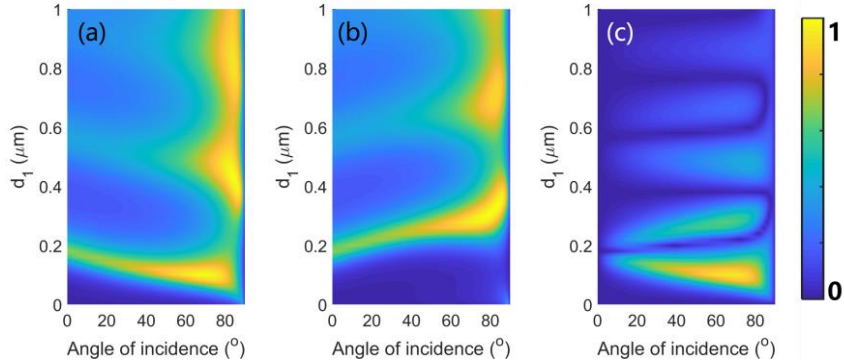

Fig. 8 (a) Absorption, (b) emission, and (c) the difference between them varying with the angle of incidence and the thickness of the MWSM at the wavelength of 9.15 μm.

## 4. Conclusions

In summary, the nonreciprocal thermal radiation based on MWSM film is investigated. The transfer matrix method is used to calculate the absorption and emission. The results show that strong nonreciprocal radiation at the wavelength of 9.15 μm can be achieved when the thickness of the MWSM film is 100 nm. The enhanced nonreciprocity is attributed to the FP resonances. Our results can not only deepen our understanding about the nonreciprocity of the MWSM films, but also show that the MWSM film is the promising candidate to engineer the ultrathin and simple nonreciprocal thermal emitters.

**Acknowledgements**

The authors acknowledge the support of the National Natural Science Foundation of China (Grant Nos. 61405217, 52106099 and 12104105)，the Zhejiang Provincial Natural Science Foundation (Grant No. LY20F050001), the Anhui Provincial Natural Science Foundation (Grant No. 2108085MF231), the Anhui Polytechnic University Research Startup Foundation (Grant No. 2020YQQ042), the Pre-research Project of National Natural Science Foundation of Anhui Polytechnic University (Grant No. Xjky02202003), the Natural Science Foundation of Shandong Province (Grant No. ZR2020LLZ004), and the Start-Up Funding of Guangdong Polytechnic Normal University (Grant No. 2021SDKYA033).